\documentclass[superscriptaddress,twocolumn,showpacs,showkeys,amsmath,prb,final]{revtex4}

\usepackage{graphicx}
\usepackage{dcolumn}
\usepackage[sort&compress]{natbib}
\usepackage{subfigure}
\usepackage{bm}
\usepackage{ifpdf}

\ifpdf
\usepackage[pdftex,
        colorlinks=true,
        pdftitle={Transverse magnetization and torque in asymmetrical mesoscopic superconductors},
        pdfauthor={Antonio R. de C. Romaguera},
        pdfsubject={Vortex theory},
        pdfkeywords={Department of Physics , Universiteit Antwerpen, Groenenborgerlaan 171 / U 208 B-2020 Antwerpen},
        baseurl={http://www.cmt.ua.ac.be},
        pdfpagemode=UseNone,
        bookmarksopen=true
        pdfoutput=1
        ]{hyperref}
\fi

\begin{document}

\title{Transverse magnetization and torque in asymmetrical mesoscopic superconductors}
\author{Antonio R. de C. Romaguera}%
\email{ton@if.ufrj.br}
\affiliation{Instituto de F\'isica, Universidade Federal do Rio de Janeiro, C. P. 68.528, 21941-972 Rio de Janeiro, Brazil}%
\affiliation{Departement Fysica, Universiteit Antwerpen, Groenenborgerlaan 171, B-2020 Antwerpen, Belgium}%
\author{Mauro M. Doria}%
\affiliation{Departement Fysica, Universiteit Antwerpen, Groenenborgerlaan 171, B-2020 Antwerpen, Belgium}%
\affiliation{Instituto de F\'isica, Universidade Federal do Rio de Janeiro, C. P. 68.528, 21941-972 Rio de Janeiro, Brazil}%
\author{F. M. Peeters}%
\affiliation{Departement Fysica, Universiteit Antwerpen, Groenenborgerlaan 171, B-2020 Antwerpen, Belgium}%
\begin{abstract}
We show that asymmetrical mesoscopic superconductors bring new insight into vortex physics where we found the remarkable coexistence of long and short vortices. We study an asymmetrical mesoscopic sphere, that lacks one of its quadrants, and obtain its three-dimensional vortex patterns by solving the Ginzburg-Landau theory. We find that the vortex patterns are asymmetric whose effects are clearly visible and detectable in the transverse magnetization and torque.
\end{abstract}

\maketitle

The small volume to surface ratio in mesoscopic superconductors makes their properties distinct from the bulk materials.\cite{dorsey00} For instance, in the absence of pinning and anisotropy, mesoscopic superconductors have an intrinsic magnetic hysteresis whereas bulk samples do not. Mesoscopic superconductors are intrinsically metastable since they can be in exited vortex states due to an energetic barrier upholding the decay to the groundstate.\cite{baelus01,deo99} Several experimental techniques have been developed to detect mesoscopic vortex states\cite{geim97,bolle99,geim00} and to reveal their new and interesting physical properties.\cite{grigorieva06,kanda04}
Here we obtain the transverse magnetization and torque that result from the \emph{asymmetrical} coexistence of vortices with different lengths
in mesoscopic superconductors. The general aspects of our results apply to irregularly shaped grains and, consequently, can also be
relevant to the understanding of inhomogeneous bulk samples with granular structure.\cite{jurelo97,howald01,ekinoa05}

In this paper
we obtain and compare the vortex patterns of two similar but distinct mesoscopic systems, namely, a \emph{full} sphere with
radius $R=6.0\xi$ and an asymmetrical sphere, with the same radius
but lacking one of its quadrants. Hereafter we call it
\emph{minus-quarter} sphere and present here its transverse
magnetization and torque.
%
%
\begin{figure}[t]
\includegraphics[width=\linewidth]{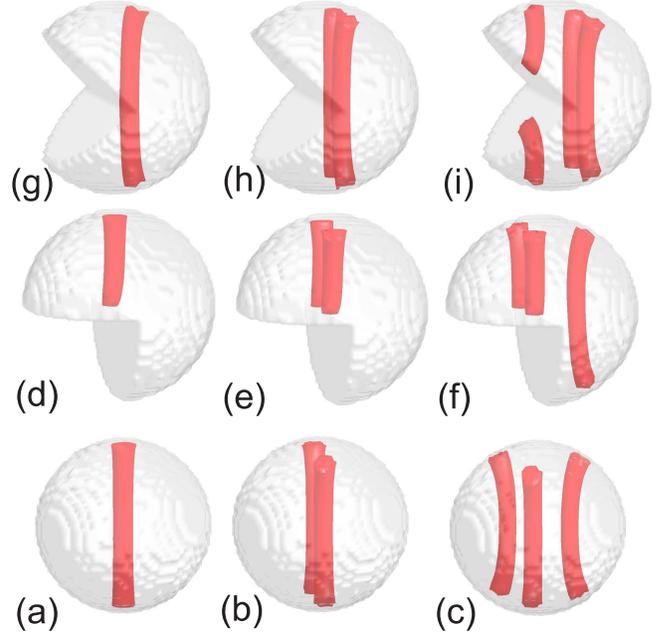}
\caption{(Color online) Three-dimensional iso-density plots taken at
20$\%$ of the maximum density $|\psi|^2$. Each iso-contour is a
single surface made of the vortices and thesuperconductor surface. The
magnetic field is directed from bottom-to-top with $H/H_{c2}=0.16$ (a), 0.29 (b), 0.35 (c), 0.27 (d), 0.39 (e),
0.49 (f), 0.35 (g), 0.48 (h), and 0.54 (i). The figures correspond to
the \emph{full }sphere (a)-(c) and to the \emph{minus-quarter}
sphere, positioned in two ways, ((d)-(f) and (g)-(i), with respect
to the field.} \label{fig1}
\end{figure}
%
%
There are many reasons for the magnetic moment of a sample with
volume $V$, ${\bm \mu}=V {\bm M}$, not to be aligned along the
applied magnetic field, ${\bm H}$. A direct measurement\cite{doria94,kogan02,wang05} of the torque
\begin{eqnarray}
{\bm \tau} = {\bm \mu} \times {\bm H}=H \mu \sin{ \theta } \bm {\hat e_{\tau}},
\label{eq1}
\end{eqnarray}
brings insight about these distinct contributions,\cite{farrel88,won96} where $\theta$ is the angle between $\bm H$ and $\bm M$. For example, vortices pinned by
point-like normal inclusions (defects) are major contributors to torque due to their inability to follow a rotating applied field. This contribution is usually avoided as one seeks out
true information about intrinsic properties of the superconducting
state. The sample's shape and size is also another contributor and
provides the basis for very high precision magnetization
measurements using torque.\cite{wang05} Basically it relies on the
thin film limit, where vortices remain straight and single oriented
inside the sample upon field rotation, resulting in a magnetic sign
always perpendicular to the major surfaces. The general misalignment
between the applied field and the sample's major axis produces a
torque, even for a regularly shaped superconductor. Hence to obtain
a torque it suffices that the circulating supercurrent yields a
magnetic field response not oriented along the applied field that
gave rise to it. Asymmetry leads to such effect and we show
here that it provides a useful tool to probe the physics of
mesoscopic superconductors.

In this paper we provide the first study of a mesoscopic
superconductor with a transverse ($M_x$) magnetization, thanks to
the use of a truly three-dimensional approach.\cite{doria07,roma07}
The found effects in the transverse magnetization and in
the torque are within reach of present experimental detection.\cite{wang05} For instance, a coherence length of $\xi \simeq 100$
nm leads to $H_{c2}=\Phi_0/2\pi \xi^2 \simeq 300$
Gauss.\cite{deo97} If the penetration length is comparable to the
sphere's radius R = 0.6 $\mu$m, then the transverse magnetization is
approximately ten times smaller, $M_x \simeq 10^{-3} H_{c2} \simeq
0,3$ Gauss than the longitudinal one $M_z \simeq 10^{-2} H_{c2}
\simeq 3$ Gauss. The magnetic moment of this sphere is $\mu \sim M_z
V \simeq 7.0 \;\textrm{x}\; 10^{-12}$ emu ($V=72\pi \xi^3 \simeq
7.63 \;\textrm{x}\; 10^{-13}$ cm$^{3}$ for the \emph{minus-quarter}
sphere). The torque, $\tau_y = -M_xH_z V \simeq 10^{-3} H_{c2}^2 V$
becomes $\tau_y = 4.0 \;\textrm{x}\; 10^{-12}$ erg.

\begin{figure}[b]
\centering
\includegraphics[width=\linewidth]{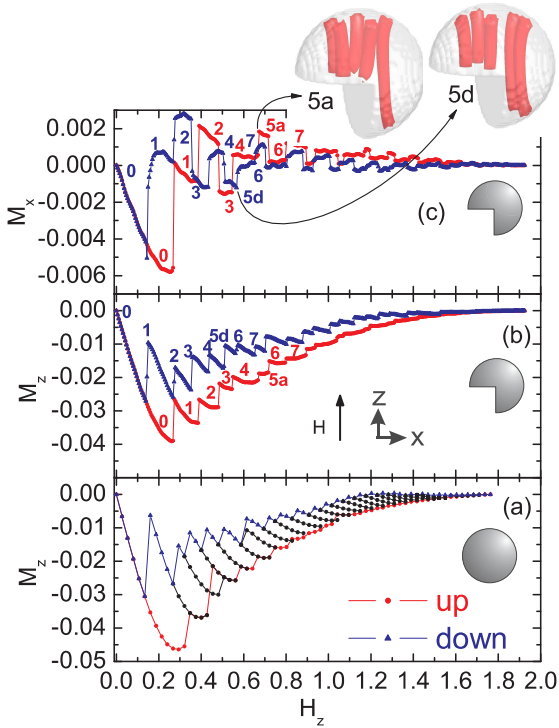}
\caption{(Color online) The magnetization for a full field cycle is
shown here, (a) for the \emph{full } and (b)-(c) for the
\emph{minus-quarter} sphere. The ascending (red) and descending
(blue) branches are labeled for the first nine vortex states.
Selected iso-plots obtained in ascending and descending fields show
the degeneracy of the 5 state. The full non-intersecting
magnetization branches are shown just for the \emph{full} sphere.}
\label{fig2}
\end{figure}

Fig.~\ref{fig1} summarizes the novel aspects brought by asymmetry showing the Cooper pair density, $|\psi(\bm r)|^2$, obtained
from our numerical simulations for selected values and orientations of the applied field. States with at most three vortices were selected
to help to understand the basic effects caused by asymmetry.
For all cases the magnetic field is oriented along the bottom-to-top
direction. In those mesoscopic systems vortices are naturally curved\cite{PhysRevB.romaguera.inpress} because the entrance and exit of the vortex lines must be perpendicular to the surface,\cite{brandt81} a fact that causes a closely packed configuration near to the equatorial region,
as previously found.\cite{doria07}

\begin{figure}[t]
\centering
\includegraphics[width=\linewidth]{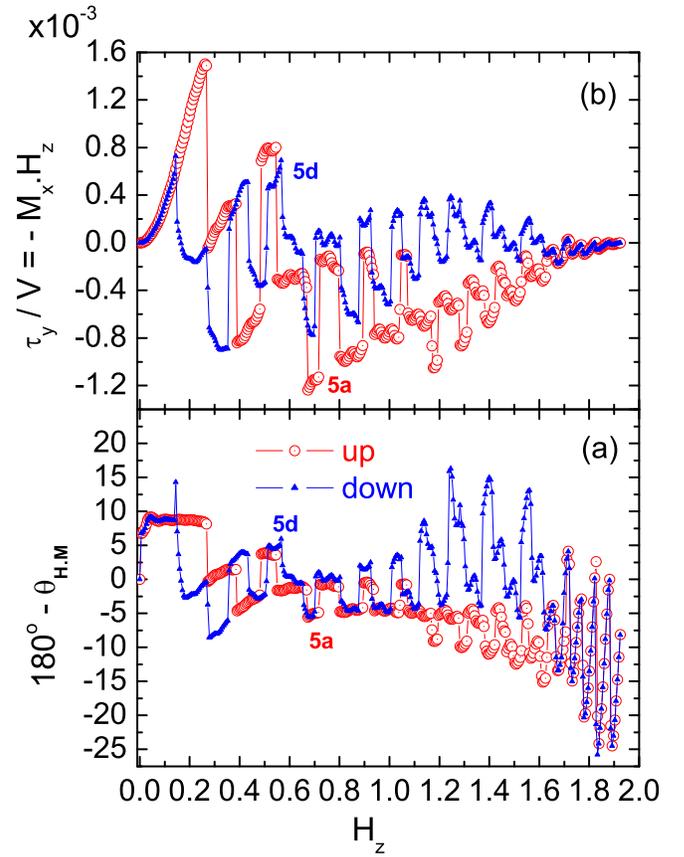}
\caption{(Color online) The ascending (red) and descending (blue)
torque branches are shown here in (a) sub-plot and the corresponding
misalignment angle in (b) sub-plot.} \label{fig3}
\end{figure}

Fig.~\ref{fig1}(d) best represents
the effects of the asymmetry where theposition of the short vortex very close to the
center of the \emph{minus-quarter} sphere is determined by: (i) the spherical shell pushing the vortex near the central axis where the shielding current offers maximal stability\cite{baelus07} and, (ii) the flat surface of the missing quadrant where the vortex length is half of the one for a full sphere. As a
consequence, the vortex is displaced from the central axis, breaking the rotational symmetry in the northern hemisphere resulting in transverse magnetization and torque.
The second vortex has the same faith of the first one, it is also a
short vortex that gets as close as possible to the central axis
(Fig.~\ref{fig1}(e)). The situation changes for the third vortex,
possibly because $R=6.0\xi$ is just too small to accommodate it in
the quadrant, forcing its nucleation in the full hemisphere where it
becomes the first long vortex. Asymmetry brings peculiar, not always
intuitive, features. For instance the rotation of the
\emph{minus-quarter} sphere by 45$^\circ$ leads to new vortex
configurations not accessible in case of the previous alignment. Now
the long vortices nucleate first and short vortices enter in pairs
afterwards, as shown in Figs.~\ref{fig1}(g)-(i), with increasing field. In summary asymmetry brings new situations that
can bring further information about vortex patterns in mesoscopic
superconductor. They are detectable in the transverse magnetization
and the torque, bringing inside information about vortex patterns in
such systems.

The Ginzburg-Landau theory, derived near the critical temperature,
is applicable at much lower temperatures for the mesoscopic
superconductors, where it was found to give a fair description of
its magnetic properties.\cite{geim97,deo97} We solve the full
three-dimensional Ginzbug-Landau theory for the asymmetrical
superconductor by the numerical procedure described in Refs.~\onlinecite{doria04,doria07}, that minimizes the free energy with
the boundary conditions included. In this approach the
superconductor shape is characterized by a step-like function in the
free energy (1 inside the superconductor and 0 outside). Here we
also include an intermediate step inside the superconductor (0.8
within a layer of thickness 0.5$\xi$) to stabilize the states for
fields near $H_{c3}$, according to Ref.~\onlinecite{doria07}. The
present results are under the approximation that the applied field
uniformly permeates the mesoscopic superconductor so that the
magnetic shielding is safely ignored. A discretized gauge invariant
version of the Ginzburg-Landau free energy is minimized in a cubic
mesh with 46 x 46 x 46 points\cite{doria07} and the magnetization
is directly determined from $\bm M = const \;(1/2c)\; \int \; \bm r
\times \bm J \; dv/V $, where $\bm J$ is the supercurrent in the
mesoscopic superconductor volume and $const$ a free parameter
associated to the demagnetization tensor\cite{osborn45,beleggia06} $\bm D$, and so, obtained in our numerical
simulations from the zero vortex branch (Meissner phase) at small
$H$: $\bm H +4\pi \bm D \bm \cdot \bm M=0$. We apply the
inverted relations $M_x = (-D_{zz}H_x+D_{xz}H_z)/det$ and $M_z =
(D_{zx}H_x-D_{xx}H_z)/det$ with $det = D_{zz}D_{xx}-D_{xz}D_{zx}$ in
our numerical analysis to find that for the \emph{full} sphere the demagnetization
tensor reduces to a single scalar, $D$, as expected, which is
adjusted to the well known value\cite{osborn45,beleggia06} of $D=1/3$
and for the \emph{minus-quarter} sphere
$D_{xx}=D_{zz}$, $D_{xz}=D_{zx}$, and also that $D_{zz}/D=1.20$ and
$D_{zx}/D=-0.184$. In the rest of the paper we use the gaussian system of units and express both
the magnetization and the applied field in terms of $H_{c2}$, the
bulk upper critical field, and the torque per volume in units of
$H_{c2}^2$.

Fig.~\ref{fig2} shows the magnetization of the \emph{full} and the
\emph{minus-quarter }spheres. We use the notation
\emph{long}+\emph{short}$\rightarrow$ (\emph{long}, \emph{short}) to
express the \emph{minus-quarter }sphere states, whose first nine
ones, labeled in Figs.~\ref{fig2}(b) and (c), are given by: $0
\rightarrow(0,0)$, $1 \rightarrow(0,1)$, $2 \rightarrow(0,2)$, $3
\rightarrow(1,2)$, $4 \rightarrow(1,3)$, $5d \rightarrow(1,4)$, $5a
\rightarrow(2,3)$, $6 \rightarrow(2,4)$, and $7 \rightarrow(2,5)$.
Notice that there are two $5$ vortex states, $5a$, and $5d$, the
first being the lowest in energy, and therefore reached by the
ascending field. These two states can coexist under the same field
although Fig.~\ref{fig2} does not show the corresponding
prolongation of the $5a$ and $5d$ curves. This degeneracy is a new
and interesting aspect brought by asymmetry which shows that the total number of vortices not always determines iniquely the magnetization branch. We found 24 states for the \emph{full} spheres
and at least 26 for the \emph{minus-quarter} sphere. The higher
vortex states, which live near to the normal state, become
increasingly difficult to identify and so, required more precision
than considered here. The magnetization curves were obtained to a
full cycle field sweep that starts at zero, reaches the maximum
possible value ($H_{c3}$) (ascending branch, red color), and
finally is lowered back to zero (descending branch, blue color). Figs.~\ref{fig2}(a) and (b) show the magnetization component
along the field, $M_z$, with the expected feature of rounded
branches for ascending field and of a tooth-saw structure for
descending field. This behavior has been extensively discussed both
theoretically\cite{schweigert98,schweigert99,baelus02} and also
experimentally.\cite{geim97}
The full stability region for the different vortex configuration is only displayed here for the
\emph{full} sphere (Fig.~\ref{fig2}(a)). Its construction requires
numerous sweeps starting at intermediate applied field values.
Notice the distinct features of $M_x$ in comparison to the $M_z$.
The ascending and descending branches are entangled and have an
underlying structure that approximately breaks in distinct families,
associated to odd and even number of vortices. These branches
display a wiggled structure due to vortex repositioning that cause
abrupt changes in vortex length, and so in energy and in current
flow.
\begin{figure}[b]
\centering
\includegraphics[width=\linewidth]{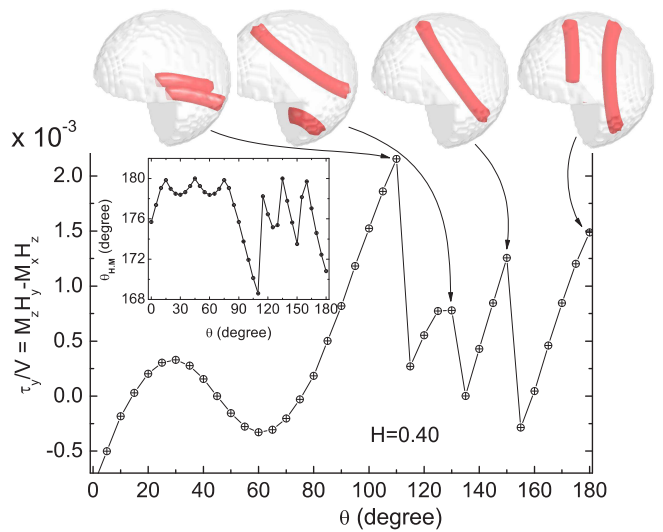}
\caption{(Color online) The torque is shown upon rotation of the
fixed applied field ($H = 0.40$) in 5$^\circ$ increments. The inset
shows the angle between ${\bm M}$ and ${\bm H}$ versus the applied
field angle. Selected iso-plot configurations for 110$^\circ$,
130$^\circ$, 150$^\circ$, and 180$^\circ$ are also displayed here.}
\label{fig4}
\end{figure}

The vortex entrance or exit can be detected through the
transverse magnetization, whose sign changes accordingly, as seen in
Fig.~\ref{fig2}(c). Mesoscopic superconductors are known to be able to exhibit a
paramagnetic signal when in a metastable vortex state\cite{palacios00,geim98} but this is not the
case for the sphere, as the dominant component ($M_z$) remains
steadily oppositely oriented to the applied field. However a
change of sign in $M_x$ implies that the torque flips direction, as
shown in Fig.~\ref{fig3}. Moreover the torque is proportional to the
applied field and it gets naturally enhanced in the regime between
$H_{c2}$ and $H_{c3}$, contrary to the magnetization, which gets dim
in this region. The magnetization angle, 180$^\circ-\theta$, shows that the
misalignment with the applied field is about 10$^\circ$ for small fields and it oscillates above $H_{c2}$. Thus, torque measurement can be very helpful in
the detection of high vortex states where the onset of a vortex configuration causes an abrupt
change of sign.
As an example, Fig.~\ref{fig4} shows the torque for a rotating
field of intensity $H=0.4$. The selected initial configuration for $0^\circ$
is the ground state for this field, made of two short vortices (see
Fig.~\ref{fig1}(e)). As the field rotates by an angular increments of
$5^\circ$, the two short vortices follow along and this
configuration remains stable until $110^\circ$, albeit the fact that
during this process the torque flips sign four times! The torque
suffers an abrupt drop at $115^\circ$ signaling the entrance of a
new vortex pattern. One of the short vortices is replaced by a long
vortex and the remaining short vortex moves to the center of the
\emph{minus-quarter} sphere. This long and short vortex configuration, i.e. $(1,1)$
lasts until the next torque transition at $130^\circ$, where a
single long vortex becomes the accessed configuration (from
$135^\circ$ until $150^\circ$). Thus the Fig.~\ref{fig1}(g) state
is reached in this way. The last abrupt drop brings back the $(1,1)$
configuration from $155^\circ$ until $180^\circ$. The final state
does not coincide with the initial one, and is not included as
one of the available stable configurations listed in Fig.~\ref{fig1}.
Thus it is a very high excited state much above the ground state.
Nevertheless it became accessible by virtue of the torque.

In conclusion we have shown here that the extra complexity of vortex
patterns in asymmetrical mesoscopic superconductors, under a tilted field, can be used as a tool for
vortex detection through the mesurement of the transverse magnetization and the
torque.

A. R. de C. Romaguera acknowledges the Brazilian agency CNPq for
financial support. M. M. Doria thanks CNPq, FAPERJ, Instituto do
Mil\^enio de Nanotecnologia (Brazil) and BOF/UA (Belgium). F. M.
Peeters acknowledges support from the Flemish Science Foundation
(FWO-Vl), the Belgian Science Policy (IUAP) and the ESF-AQDJJ
network.
\bibliography{torque}
\end{document}